\documentstyle[11pt,epsfig]{elsart}
\begin{document}
Accepted for publication in Physics Letters B\\
%, version \today\\
\begin{frontmatter}
\title
{Transition from Thermal to Rapid Expansion
%{Transition from Thermal Break-up to Dynamic Disintegration
in Multifragmentation of Gold Induced by Light Relativistic Projectiles}
%----------------------------------------------------------------
%---- Author's address                                       ----
%----------------------------------------------------------------
% Helmut Oeschler\\
%Institut f\"ur Kernphysik
%Technische Universit\"at Darmstadt
%Schlossgartenstr. 9
%D-64289~Darmstadt
%
% FAX:   +49-6159-71-.... (GSI)\\
% Phone: +49-6151-16-2221 (TU)\\
% Phone: +49-6159-71-2612 (GSI)\\
%----------------------------------------------------------------
%---- Author and coauthors                                   ----
%----------------------------------------------------------------
\author[Dub]{S.P.~Avdeyev},
\author[Dub]{ V.A.~Karnaukhov},
\author[Dub]{L.A.~Petrov},
\author[Dub]{V.K.~Rodionov},
\author[Dub]{V.D.~Toneev},
\author[TUD]{H.~Oeschler},
\author[Kur]{O.V.~Bochkarev},
\author[Kur]{L.V.~Chulkov},
\author[Kur]{E.A.~Kuzmin},
\author[Cracow]{A.~Budzanowski},
\author[Cracow]{W.~Karcz},
\author[Cracow]{M.~Janicki},
\author[IOWA]{E.~Norbeck},
\author[Bot]{A.S.~Botvina}
\address[Dub]{Joint Institute for Nuclear Research, 141980 Dubna, Russia}
\address[TUD]{Institut f\"ur Kernphysik,
Technische Universit\"at Darmstadt, D-64289~Darmstadt, Germany}
\address[Kur]{Kurchatov Institute, 123182, Moscow, Russia}
\address[Cracow]{H.~Niewodniczanski Institute of Nuclear Physics, 
31-342, Cracow, Poland}
\address[IOWA]{University of Iowa, Iowa City, IA 52242 USA}
\address[Bot]{Institute for Nuclear Research, 117312, Moscow, Russia}

\vspace*{.1cm}
%----------------------------------------------------------------
%---- Abstract                                               ----
%----------------------------------------------------------------
\begin{abstract}
 Multiple emission of intermediate-mass fragments has been
studied for the collisions of p, $^4$He and $^{12}$C on Au
with the $4\pi$ setup FASA.  
In the case of $^{12}$C(22.4 GeV)+Au and
 $^4$He(14.6 GeV)+Au collisions, the  deviations from a pure
thermal break-up are seen in the  energy spectra of the emitted
fragments: the spectra are harder than calculated and than  
measured in p-induced collisions. This difference  is attributed
to a collective flow with the expansion velocity on  
the surface about 0.1~$c$ (for $^{12}$C+Au collisions).  
\end{abstract}
\end{frontmatter}

\newpage                  

Nuclear multifragmentation is a new decay mode of excited nuclei
characterized by the emission of
Intermediate Mass Fragments (IMF, $3 \leq Z \leq 20$).
The development of this field has been strongly stimulated by the idea that this
process might be related to a liquid-gas phase transition in nuclear matter.
A recent status on multifragmentation can be found in Ref.~\cite{5}. 

Highly excited nuclei can be formed by
heavy ion collisions at intermediate energies, but there are advantages
in producing the excitation by
light relativistic projectiles. In the first case,
nuclear heating  is accompanied by
compression, fast rotation and shape distortion which may cause
dynamic effects in the multi-fragment disintegration and
it is not easy to
disentangle all these effects and extract information on  thermodynamic
properties of hot nuclear systems. 
The situation becomes
more transparent if light relativistic projectiles are
used.  In this case,  dynamic effects are expected to be negligible.  Another
advantage is that all the fragments are emitted by a single source: a slowly
moving target remainder. Its excitation energy might be
almost entirely thermal.
Light relativistic projectiles provide therefore an unique possibility to study
``thermal multifragmentation''.
It has been shown that thermal  multifragmentation indeed takes place
in collisions of light relativistic projectiles (p, $^3$He,
 $\alpha$ and recently $\pi$ and $\bar {\rm p}$)
with a heavy target 
\cite{Yen,6,7,8,9,10,11,Goldenbaum,Fox,12,Lefort}.

In this Letter we concentrate on studying energy spectra of the
emitted fragments as they reflect, due to the ``Coulomb law'', the geometry
and dynamics (expansion) of the emitting source.
By comparing our results from p+Au~\cite{7} collision with those
from reactions induced
by $\alpha$ particles and $^{12}$C projectiles with incident energies
of (1 -- 4) $A\cdot$GeV,
we evidence a transition from a pure statistical process
(``thermal multifragmentation'') to a behaviour reflecting dynamics.
It will be shown that from the observed additional collective energy
a spatial distribution of the fragments at freeze out can be infered.

The experiments were performed with beams from the JINR
synchrophasotron in Dubna using the modified \cite{13}
$4 \pi$-setup FASA \cite{14}.
The device consists of two main parts :
(i) Five  $\Delta E$ (ionization chamber) $\times$
$E$ (Si)-telescopes, which serve as a trigger for the read-out of the system
allowing measurement of the charge and energy distributions of IMF's at
various angles from $24^\circ$ to $156^\circ$
covering together a solid angle of 0.03 sr. (ii) The fragment multiplicity
detector consisting of 64 CsI(Tl) counters
(with thicknesses around 30 mg$\cdot$cm$^{-2}$) which covers 89\% of $4 \pi$.
This device gives the number of IMF's in the event and 
their spatial distribution.
A self-supporting Au target of 1.5 mg/cm$^2$ thickness was located in the center
of the FASA  vacuum chamber ($\sim 1$~m in diameter).
The following beams were used: protons at energies of 2.16, 3.6 and 8.1 GeV,
$^{4}$He at energies of 4 and 14.6 GeV and $^{12}$C at 22.4 GeV.
The average beam intensity was $7 \cdot 10^8$ particles/spill for protons
and helium and $1 \cdot 10^8$ particles/spill for carbon
projectiles with a spill length at 300 ms and a spill period of 10 s.
See also \cite{7} reporting on the p+Au experiment.

The kinetic energy of fragments is determined
by four terms: thermal motion, Coulomb repulsion,
rotation, and  collective  expansion energies  of the system at freeze
out, $E = E_{th} + E_C +E_{rot} + E_{flow}$.  The additivity of the first three
terms is quite obvious.  For the last term, its independence  of others
may be
considered only approximately when the evolution
of the system  after the freeze-out point is driven only by the Coulomb force.
The Coulomb term is significantly larger than the thermal one as was shown in
Ref.~\cite{10} for  $^4$He (14.6 GeV)+Au collisions, where the Coulomb part of
the mean energy of the carbon fragment is three times
larger than thermal one  
using volume emission of fragments from
a diluted system.

The contribution of the collective flow for the p+Au collisions
at 8.1 GeV incident energy was estimated to $v_{flow} < 0.02~c$
in Ref.~\cite{7}.
This was done by comparing  the measured IMF spectra with the
ones calculated within the framework of the Statistical Multifragmentation
Model (SMM)~\cite{16} which  includes no  flow.
For heavy ion collisions,
collective flow has been observed and it is the most pronounced in
central Au+Au collisions ~\cite{18}.
In this respect, it is interesting to analyse energy
spectra of fragments from He+Au and C+Au collisions looking for a possible
onset of collective flow phenomenon.

\begin{figure}
\epsfig{width=11cm,file=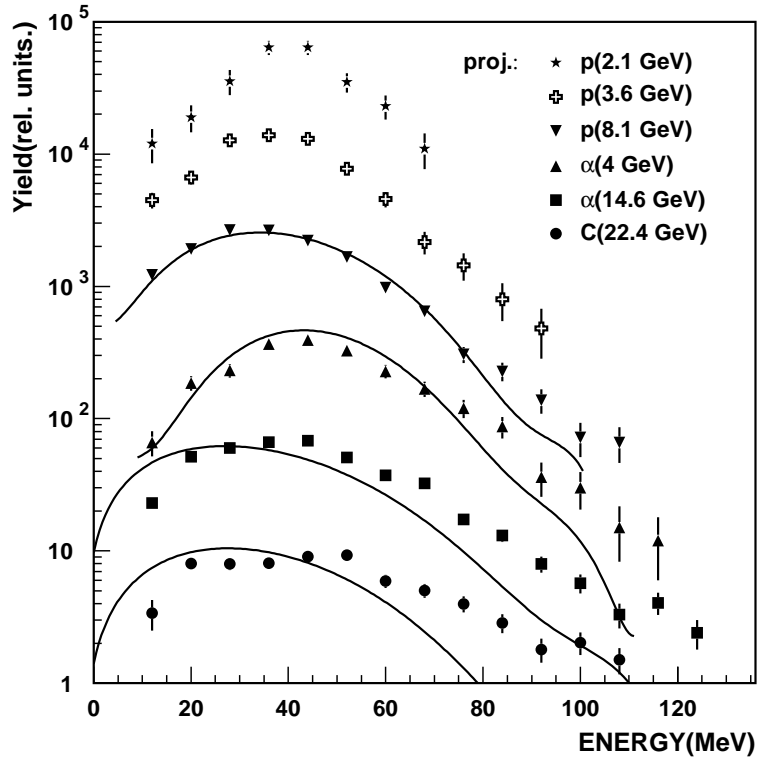}
\caption{Energy distribution of carbon fragments obtained  for
        different collision systems at $\theta = 89^\circ$.
        The lines are calculated in the INC+Expansion+SMM model assuming
        no flow.  The spectra are arbitrarily scaled along the yield axis
        to avoid mixing of symbols.}
\label{E_carbon}             
\end{figure}

A comparison of the energy spectra of outgoing carbon isotopes from
proton-, helium- and carbon-induced collisions on a Au target
is presented in Fig.~\ref{E_carbon}.
The spectral shapes show an increase in the number of high-energy carbon
fragments with the projectile mass.

The reaction mechanism for light relativistic
projectiles is usually divided into two stages. The first one is a
fast energy-depositing stage, during which very energetic light
particles are emitted and a nuclear remnant (target spectator) is excited.
We use a refined version of the intranuclear cascade model (INC)~\cite{Toneev}
for describing the first stage.
The second stage is described by the 
Statistical Multifragmentation Model (SMM), which
considers multibody decay (volume emission) of a hot expanded nucleus.
But such a two-stage approach fails to describe the observed IMF 
multiplicities  as shown in Table 1.
An expansion stage is inserted between the two parts of 
calculation. The excitation energies and the residual
masses are then fine tuned~\cite{7}  to get an agreement 
with the measured IMF multiplicities, i.e.~the values for the residual
(after INC) 
masses $A_R$ and their excitation energies $E_R$ are scaled on 
an event-by-event basis. The average masses of nuclei which decay by
multifragmentation is labelled $A_{MF}$ having mean excitation
energies $E_{MF}$ as given in Table~1 together with the values of $<M_{IMF}>$.
The lines in Fig.~\ref{E_carbon} give the spectra calculated in the framework 
of this combined model INC+Expansion+SMM. The fragment energies are obtained 
by the multibody Coulomb trajectory calculations on an event-by-event basis. 
In the initial state all the charged particles are assumed to have a thermal 
velocity only (no flow). 

The calculated carbon spectrum for p+Au collisions (at 8.1 GeV) is 
consistent with the measured one. A similar situation occurs with
$^4$He+Au collisions at 4~GeV, but not with  
$^4$He(14.6 GeV)+Au and $^{12}$C+Au interactions: the measured spectra
are harder than the calculated ones.
A careful inspection of the reaction induced with $\alpha$ particles
shows that the maximum of the energy spectra obtained at 4 GeV is
higher than obtained at 14.6 GeV. This is mainly caused by the higher
$Z$ values of the residual nuclei at the lower incident energy and is
reproduced by the model calculations. 
The slope of the 14.6 GeV data is much harder than for the lower incicent 
energy.  This higher slope is underestimated by the model and this failure is
most pronounced for C + Au collisions.
An attempt to describe these spectra by a higher Coulomb field,
caused e.g.~by a more compact system at break-up fails. In such an
approach the maximum shifts keeping the drop towards higher kinetic
energies while the measured spectra show an increase at the higher kinetic
energies without moving the maximum significantly.

\begin{figure}
\epsfig{width=11cm,file=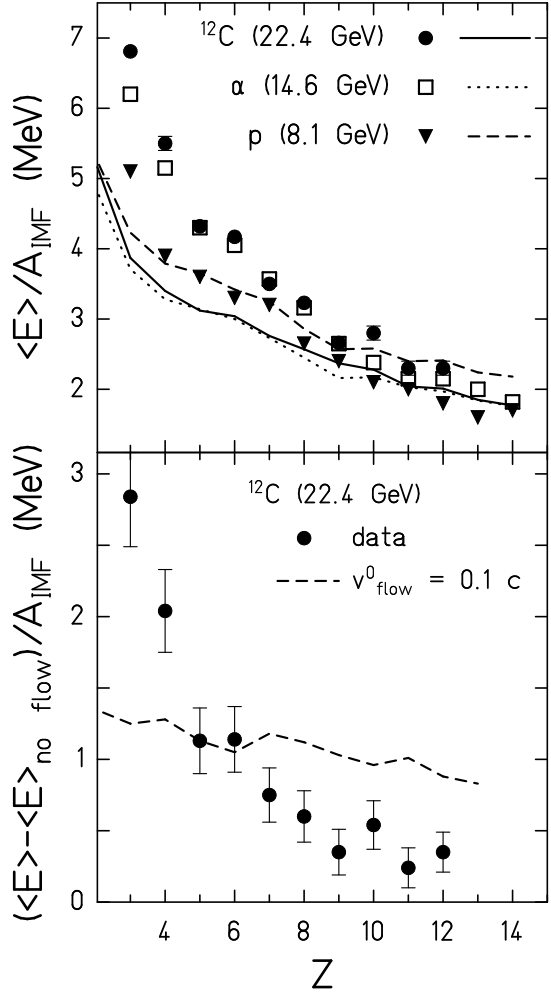}
\caption{
Upper part: The mean kinetic energies of outgoing fragments per nucleon 
measured at $\theta = 89^\circ$ for
p(8.1 GeV), $^4$He (14.6 GeV) and $^{12}$C (22.4 GeV) collisions with Au. 
The shown error bars are due to statistics only, a systematic error of 5\% 
has to be added.
The lines are calculated using INC+Expansion+SMM approach assuming no flow.
Lower part: Flow energy per nucleon (solid points) obtained as a difference
of the measured
fragment kinetic energies and the values calculated 
without flow. The dashed line shows
the calculations (see text) assuming a linear radial profile of the
expansion velocity 
with  $v^0_{flow} = 0.1~c$ (at the surface).
}
\label{E_mean}                
\end{figure}

The trends from Fig.~\ref{E_carbon}, i.e.~increasing mean energies with
increasing mass and increasing 
energy of the projectile, is seen for many emitted
fragments. This observation is summarized in Fig.~\ref{E_mean}, which shows
the mean kinetic energies per fragment nucleon
as a function of the charge of the detected fragments $Z$.
This figure shows a remarkable enhancement  in the kinetic energies for
light fragments emitted in He+Au and C+Au collisions as compared to the
p(8.1 GeV)+Au case. The calculated values of the mean fragment energies
(shown by lines) are obtained with INC+Expansion+SMM model.
The measured energies are close to the calculated
ones for
p+Au collisions in the range of the fragment charges between 4 and 9.
The experimental values for $^4$He+Au and $^{12}$C+Au interactions, however,
exceed the calculated ones, which are similar for all three cases.

The observed deviation cannot be attributed to an angular momentum effect.
The rotational energy $E_{rot}$ of a fragment with mass $A_{IMF}$ can be
estimated from the total rotational energy $E_L$ of a system
with mass  number $A_R$ using classical rotation: 
\begin{equation}
<E_{rot}> / A_{IMF} = {5 \over 3} < {E_L \over A_R} > {<R_Z^2>\over
R_{sys}^2}
\end{equation}

where $R_Z$ is the radial coordinate of the IMF and $R_{sys}$ is
the radius of the system.
According to the INC calculations for C+Au collisions, the mean
angular momentum $L$ of the target spectator  is $\approx 36 \hbar$.
It might be reduced by a factor of 1.5  due to the mass loss 
during expansion along
the way to the
freeze-out point. Finally $<E_L>$ is estimated to be only
5 MeV and $<E_{rot}>/A_{IMF} \approx 0.04$~MeV/nucleon,
which is by an order
of magnitude smaller than the energy enhancement for light fragments.
We suggest that the observed enhancement is caused by the expansion
of the system, which seems to be radial, as a
$v_{\parallel}$-versus-$v_{\perp}$ plot
(this will be subject of a forthcoming publication) 
does not show any significant deviation from circular symmetry.

An estimate of the fragment flow energy may be obtained from the difference
between the measured IMF energies $<E>$ 
and those calculated (without any flow) $<E>_{no flow}$.
This difference for C+Au collisions is shown
in the lower part of Fig.~\ref{E_mean}.
A drastic decrease of the flow energy with increasing charge of the
fragment is seen.

In an attempt to describe the data, we modified the SMM code
by including a radial velocity boost for
each  particle at freeze out, i.e. a radial expansion
velocity was superimposed on the thermal motion in the calculation
of the multibody Coulomb trajectories on an event-by-event basis.
A self-similar radial expansion of a spherical nucleus
is assumed, where the local flow
velocity is linearly dependent on the distance of the particle
from the centre of mass. The expansion velocity of particle
$Z$ located at radius $R_Z$ is given by the following expression
%8
\begin{equation}
\vec{v}_{flow} (Z) = v_{flow}^0 \cdot \frac{\vec{R}_Z}{R_{sys}}
\end{equation}
\noindent where $v^0_{flow}$ is the radial velocity on the
surface of the system.
The use of the linear profile of the radial velocity is motivated
by the hydrodynamic models for an expanding hot nuclear system
(see for example Ref.~\cite{17}).
The value of $v^0_{flow}$ has been adjusted to 0.1 $c$ in order to describe
the mean kinetic energy measured for the carbon fragments. The results are
presented in the lower part of Fig.~\ref{E_mean}
as dashed line calculated from the difference of the fragment energies
obtained for $v^0_{flow}$=0.1 $c$  and  $v^0_{flow}$=0.
The data deviate significantly from the calculated values for Li and Be.
This may be caused in part by the contribution of particle emission
during the early stage of expansion from the hotter and denser system.
It is supported by the fact that the extra energy of Li fragments
with respect to the calculated value is
clearly seen in Fig.~\ref{E_mean}
even for the proton-induced fragmentation, where no
significant flow is expected. This peculiarity of  light fragments
has been noted already by the ISIS group for   $^3$He+Au   collisions
at 4.8 GeV \cite{Fox}.

For fragments heavier than carbon, the calculated curve in the lower part
of Fig.~\ref{E_mean} is higher than the data and only slightly decreasing with
increasing fragment charge.
The trend of the calculation is to be expected.
The mean fragment flow energy
is proportional to $<R_Z^2>$ and this value changes only little with
fragment charge in the SMM code due to the assumed equal probability
of IMF distribution inside the available break-up volume.
The difference between data and calculations shown in Fig.~\ref{E_mean}
indicates that a uniform density distribution is not fullfilled.
A preference of finding heavier fragments more in the center
would reduce their Coulomb energies shown in the upper part of 
Fig.~\ref{E_mean}, and would increase the extracted quantity
$<E^{measured}>-<E>_{no flow}$ leading then to a consistent description
of a flow scenario with a non-uniform fragment distribution.
The dense interior of the expanded nucleus may be
favored for the appearance of larger IMF's,
 if  fragments are formed via the density fluctuations.
Indications for such an effect could already be drawn from the
analysis of the mean IMF energies performed in Ref.~\cite{7}
for proton induced fragmentation:
the measured energies are below the theoretical
curve for fragments heavier than Ne.
This means that our procedure of estimating the flow energy is only
reliable for fragments lighter than Ne where the model calculations
fit the data of the fragment energies from p+Au collisions (no flow).
For heavier fragments the flow energies are underestimated and should be 
considered as lower limits.
The interesting feature of a reduction of the flow energy for heavier fragments
is observed also for the central heavy ion collisions ~\cite{21}.
This effect is increasingly important at energies $\leq$100 $A\cdot$MeV, 
and that is in accordance with our suggestion on its relation to the density
profile of the hot system at freeze out.

\begin{figure}
\epsfig{width=11cm,file=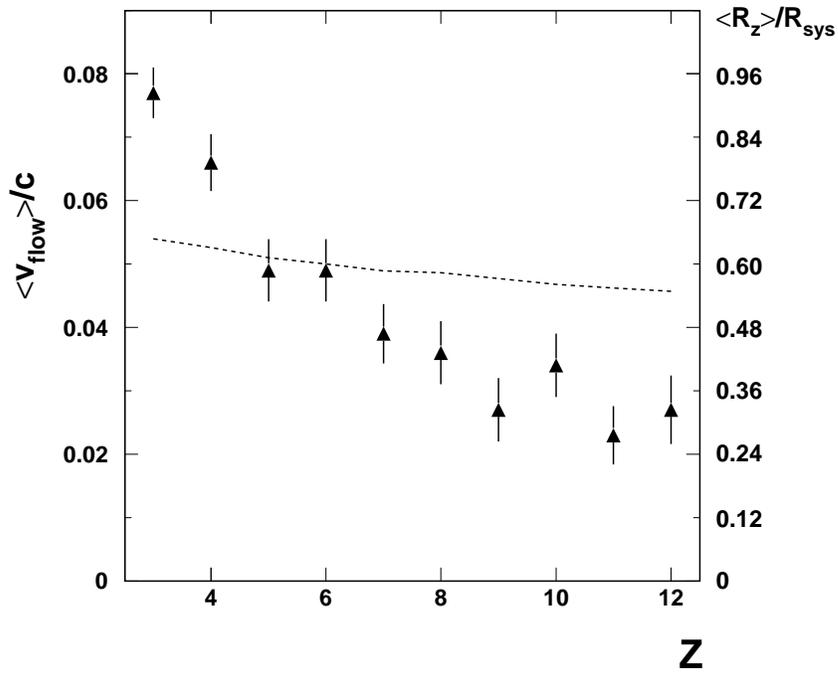}
\caption{
Experimentally deduced mean flow velocities $<v_{flow}>/c$ (left scale) 
for $^{12}$C+Au collisions  (triangles) 
as a function of the fragment charge, 
and the mean relative radial coordinates of fragments $<R_Z>/R_{sys}$
(right scale), obtained 
under the assumption of a linear radial profile of the expansion velocity.
}
\label{R_Z}
\end{figure}

For the estimation of the mean fragment flow velocities $<v_{flow}>$
the difference
between the measured IMF energies and calculated ones (without flow)
has been used. The results are presented in Fig.~\ref{R_Z}.
The values for Li and  Be are considered as upper limits because
of the possible contribution of the preequilibrium emission.
The corresponding values of the fragment positions 
$<R_Z>/R_{sys}$ for the freeze-out point 
obtained under assumptions of
the linear radial profile for the expansion velocity are given on the
right-hand scale of this figure. 
The dotted line shows
the mean radial coordinates of fragments according to the SMM code. As it has
been noted above, the calculated values of $<R_Z>/R_{sys}$ are only slightly
decreasing with $Z$ as expected from a uniform density distribution,
but in clear contrast to the data. Calculations with the MMMC 
model~\cite{Gross} give the same trend as the SMM model~\cite{Lauret}.

Effects of the radial collective energy for 1 $A\cdot$GeV Au+C collisions
(in inverse kinematics) were considered in \cite{19} by analysing 
the transverse
kinetic energies. The mean radial flow velocities were estimated, but 
it had been done only
for fragments with $Z\leq 7$. In this charge range our analysis gives
slightly lower values of the mean expansion velocities as compared to 
Ref.~\cite{19}. Their interpretation of a time sequence of the
emission acccording to the $Z$ value is largely equivalent to our
interpretation. 

The total expansion energy can be estimated by integrating the nucleon flow
energy (taken according to  Eq.~(2)) over the
available volume at freeze out. For an uniform system one obtains
%9
\begin{equation}
E_{flow}^{tot} = 
{3 \over 10} A_R \cdot m_N \ (v^0_{flow})^2 \ (1- r_N/R_{sys})^5
\end{equation}
\noindent with the mass $m_N$ and the radius $r_N$ of a nucleon.
For $^{12}$C+Au collisions it gives $E_{flow}^{tot} \simeq 100$~MeV,
corresponding to the flow  velocity at the surface of $ 0.1c$.
Similar results are obtained for $^{4}$He(14.6 GeV)+Au collisions.
The total flow energy of the fragmenting nucleus
is four times less than the thermal one estimated in 
the INC+Expansion+SMM approach.

Concluding, the energy spectra of IMF's for reactions  p+Au at 2.1, 3.6 and
8.1 GeV, $^4$He+Au at 4 and 14.6 GeV, $^{12}$C+Au at 22.4 GeV are compared.
While the fragment kinetic energies
are well described within the SMM code  for p+Au collisions, the model
underestimates the kinetic energies of fragments from collisions induced
by $^{4}$He (14.6 GeV) and $^{12}$C (22.4 GeV) projectiles.
The additional energy is attributed to collective expansion.
However, a linear flow profile fails to describe the variation of
flow energies extracted from the measured spectra with the fragment charge.
This discrepancy might be caused by the fact that
the model assumes a uniform density distribution and, hence,
a rather uniform probability distribution of forming fragments
in the freeze-out volume. The data indicate that heavy fragments are
preferentially  located more in the interior of the nucleus.

The presented study shows that in spite of the success of statistical
multifragmentation models in describing the partitions,
the freeze-out condition
might be still too simplified.
The energy spectra provide sensitive probes for the source configuration
and the emission dynamics.
The range of projectiles, from protons
to light nuclei, seems to be quite attractive in this respect
for showing a transition from ``thermal break-up'' to a 
disintegration possibly caused by rapide expansion likely together with a 
non-uniform density profile of the excited nucleus.

  The authors are thankful to Profs. A.~Hrynkiewicz, A.M.~Baldin,
S.T.~Belyaev, A.I.~Malakhov
and N.A.~Russakovich for support. 
The research was supported in part by Grant
No 00-02-16608 from Russian Foundation for Basic Research, by Grant 
No 2P03 12615 from the Polish State Committe for Scientific Research, 
by Grant No 94-2249 from INTAS, by 
Contract No 06DA819 with Bundesministerium
f\"ur Forschung und Technologie, 
by Grant PST.CLG.976861 from NATO,
and by US National Science Foundation.

\begin{center}
\begin{table}
\begin{tabular}{|c|c|c|c|c|c|c|c|c|}  \hline
$E_{inc}$& Proj  & Exper.     &\multicolumn{5}{|c|}{Calculations} & Model  \\
\cline{4-8}
(GeV) & &$M_{IMF}$&$M_{IMF}$ & $A_R$  & $A_{MF}$
  & $E_R$  & $E_{MF}$ &     \\
\hline
         2.16 & p&1.7$\pm$0.2 &      1.82 & 189 & 185 & 310 & 589  & INC+SMM\\ 
&      &         & 1.69 & 188 &183  & 288 & 564  & with expansion  \\ 
\hline
3.6  & p & 1.9$\pm$0.2  & 2.52  & 187 & 181 &    371 & 676   & INC+SMM   \\ 
&     &    & 1.89 & 184  &175  & 282 & 568 &  with expansion\\ 
\hline 
8.1 & p  & 2.1$\pm$0.2& 3.58 & 183 &  175 & 488 & 808   & INC+SMM\\ 
& &  & 2.0  & 176 & 158 & 259 & 529 &  with expansion  \\ 
\hline 
4.0& $^4$He & 1.7$\pm$0.2& 3.89 & 184 & 177 & 484 & 836 &   INC+SMM   \\ 
& &  & 1.77 & 177 & 161 & 238 & 502  & with expansion  \\ 
\hline 
14.6& $^4$He & 2.2$\pm$0.2&4.47 & 173 & 159 & 723 & 1132   & INC+SMM   \\  
& &  & 2.19  & 154 & 103 & 183 & 404 &  with expansion   \\ 
\hline 
22.4 &$^{12}$C&2.2$\pm$0.3&4.04 & 163&153 & 924 & 1216 &   INC+SMM   \\  
& &  & 2.17  & 139 & 86 &207 & 415  & with expansion \\ 
\hline 
\end{tabular}
\vspace*{.2cm}
\caption{
 The calculated properties of  nuclear remnants from $proj$ +  Au
collisions for INC+SMM and INC+Expansion+SMM. 
The  $M_{IMF}$ is the mean number of IMF's for events with at least
one IMF and  $A_R$, $E_R$ are the mean mass number and excitation
energy (in MeV), respectively,  averaged over all inelastic collisions, while
the quantities $A_{MF}$, $E_{MF}$ are  averaged only over residues
decaying by IMF emission.}
\end{table}
\end{center}
\newpage

%Fig.~1.

\newpage

%Fig.2. 

%Fig.~3.  

\end{document}